\documentclass[onecolumn,amsmath,amssymb,prl,superscriptaddress]{revtex4-1}
\usepackage[pdftex]{graphicx}
\usepackage{dcolumn} 
\usepackage{bm}              
\usepackage{amsmath}
\usepackage{braket}
\begin{document}
\title{Quantitative Molecular Orbital Energies within a $G_0W_0$ Approximation}

\author{Sahar Sharifzadeh}
\affiliation{Molecular Foundry, Lawrence Berkeley National Laboratory}

\author{Isaac Tamblyn} 
\affiliation{Molecular Foundry, Lawrence Berkeley National Laboratory}
\affiliation{Physical and Life Sciences Directorate, Chemical Sciences Division, Lawrence Livermore National Laboratory}

\author{Peter Doak} 
\affiliation{Molecular Foundry, Lawrence Berkeley National Laboratory}

\author{Pierre Darancet} 
\affiliation{Molecular Foundry, Lawrence Berkeley National Laboratory}

\author{Jeffrey B. Neaton} 
	\email{jbneaton@lbl.gov}
	\affiliation{Molecular Foundry, Lawrence Berkeley National Laboratory}

\begin{abstract}
Using many-body perturbation theory within the $G_0W_0$ approximation, we explore routes for computing the ionization potential (IP), electron affinity (EA), and fundamental
gap of three gas-phase molecules -- benzene, thiophene, and (1,4) diamino-benzene -- and compare with experiments. We examine the dependence
of the IP on the number of unoccupied states used to build the dielectric function and the self energy, as well as the dielectric
function plane-wave cutoff. We find that with an effective completion strategy for approximating the unoccupied subspace, and a
converged dielectric function kinetic energy cutoff, the computed IPs and EAs are in excellent quantitative agreement with available experiment (within $0.2$ eV), indicating that a one-shot $G_0W_0$ approach can be very accurate for calculating addition/removal energies of small organic molecules. Our results indicate that a sufficient dielectric function kinetic energy cutoff may be the limiting step for a wide application of $G_0W_0$ to larger organic systems.
\end{abstract} 
\maketitle
\section{Introduction}
\label{intro}

Organic molecules and assemblies are of considerable interest for next-generation photovoltaics~\cite{scharber_opv,Lloyd_opv,bredas_opv} and other energy conversion 
applications~\cite{Bard_organic,Vardeny2005}. Their performance and utility hinges on understanding and control of their spectroscopic properties, such as ionization 
potentials (IPs) in gas-phase and solid-state environments, and orbital energy level alignment at interfaces. Density functional theory (DFT) is a widely used computational 
framework for studying structural and electronic properties of materials. However, Kohn-Sham frontier orbital energies and energy differences within common approximations to DFT, 
such as the local density approximation (LDA) and generalized gradient approximations (GGAs), are known to dramatically underestimate these 
quantities~\cite{koerzdoerfer_when_2009,koerzdoerfer_erratum:_2010,Kummel_review}. Recently, we have shown that accurate fundamental gaps for gas-phase and 
solid-state organic molecules~\cite{sharifzadeh_2012}, and frontier orbital energies for an organic/metal interface~\cite{tamblyn_electronic_2011} 
((1,4) diamino-benzene on Au(111)) may be computed with many-body perturbation theory within the GW approximation~\cite{hybertsen_electron_1986}. 
For the latter, we found that the calculation must be adequately converged with respect to addition/removal energies of the \emph{isolated} components, \textit{i.e.} 
molecule and substrate. In this article, building on prior work~\cite{faber_first-principles_2011,blase_first-principles_2011,blase_charge-transfer_2011,rostgaard_fully_2010,qian_photoelectron_2011,samsonidze_simple_2011,tamblyn_electronic_2011,umari_2009,umari_gw_2010,blase_charge-transfer_2011,strange_2011}, 
we explore the extent to which we may obtain accurate IPs and electron affinities (EAs) of gas-phase molecules using a $G_0W_0$ approximation.

While there are numerous studies benchmarking the $GW$ approximation against transport gaps in bulk inorganic 
solids~\cite{onida_electronic_2002,hybertsen_electron_1986,garcia-gonzalez_many-body_2002,garcia-gonzalez_self-consistent_2001,aulbur_quasiparticle_1999}, 
similar works for isolated molecular systems are less common, and while all works exhibit marked improvement over standard DFT approaches, 
there is some quantitative disagreement (see \emph{e.g.},~\cite{faber_first-principles_2011,blase_first-principles_2011,blase_charge-transfer_2011,rostgaard_fully_2010,qian_photoelectron_2011,samsonidze_simple_2011,tamblyn_electronic_2011,kronik_photoemission_2003,umari_2009,umari_gw_2010}). 
For example, for gas-phase molecules, using an atom-centered basis set, it has been found that self-consistency in either the $GW$ eigenvalues~\cite{faber_first-principles_2011,blase_first-principles_2011,blase_charge-transfer_2011} or in both the eigenvalues and eigenvectors~\cite{rostgaard_fully_2010} 
is essential for obtaining good agreement of computed molecular IP and EA with experiment. On the other hand, with a 
planewave basis set, it has been demonstrated that a more systematic representation of the dielectric matrix and Coulomb-hole (CH) term, $\Sigma_{\textrm{CH}}$,  brings the  $G_0W_0$-predicted IP and EA in closer agreement with  experiment~\cite{qian_photoelectron_2011,samsonidze_simple_2011,tamblyn_electronic_2011}.
 Beyond the differences in their basis sets, these studies have differed in their representation of the dielectric matrix, the presence of a truncation scheme for the Coulomb interaction, and their approach for handling the empty states necessary to converge $\Sigma_{\textrm{CH}}$. 
As a consequence, the accuracy of different $GW$ approaches for the IP of gas-phase molecules remains an open question. 

Here, we compute the $G_0W_0$ IP, EA, and fundamental gap (IP - EA) of three gas-phase molecules benzene (BEN), (1,4) diamino-benzene (BDA), and thiophene (TP), as shown in 
Fig.~\ref{f:IE_convergence_epsilon}, and compare the computed IP and EA with measurements~\cite{lias_ionization_2012,cabelli_upe_1981,streets_mesomeric_1972,ben_ea}. 
We examine the dependence of the IP and fundamental gap on the number of unoccupied states used to build the dielectric function and the self energy, as well as the
 dielectric function~$\mathbf{G}$-space cutoff. We find that as our calculations approach convergence, the computed IPs and EAs are in excellent quantitative agreement 
with experiment (within $0.2$ eV), indicating that $G_0W_0$ can be very accurate for calculating addition/removal energies of small organic molecules.

\section{Methods}

Our $GW$ calculations are performed using the BerkeleyGW~\cite{deslippe_berkeleygw:_2011} package, following an established $G_0W_0$ approach~\cite{hybertsen_electron_1986}. 
The self-energy, $\Sigma=iGW$, is computed as a first order correction to the Kohn-Sham DFT Hamiltonian. The quasiparticle states are taken from DFT within the GGA of 
Perdew, Burke, and Ernzerhof (PBE)~\cite{perdew_generalized_1996} and are expanded in a planewave basis set. The cutoff for the planewave expansion is $80$ Ry for BEN and 
TP and $60$ Ry for BDA, and is determined such that the DFT total energy is converged to $<$ 1 meV/atom. The molecular geometry is optimized such that forces are less 
than $.04$ eV/$\AA$. Norm conserving pseudopotentials are used, with 1, 4, 5, and 6 electrons explicitly treated as valence for H, C, N, and S, respectively. 

Since periodic boundary conditions are imposed in our planewave DFT and subsequent GW calculations, the molecules are placed in a large supercell, chosen to be twice 
the size necessary to contain $\ge 99\%$ of their charge density. The supercell dimensions are $14$ x $8$ x $15$ $\AA^{3}$ for BEN, $14$ x $9$ x $14$ $\AA^{3}$ for TP, 
and $15$ x $15$ x $15$ $\AA^{3}$ for BDA. In constructing the dielectric matrix and the self-energy at the G$_0W_0$ step, the Coulomb potential is truncated at half 
of the unit cell length in order to avoid spurious interactions between periodic images. The electrostatic potential at the surface of the supercell is computed at the 
DFT level and its average subtracted from the GW eigenvalues to obtain absolute energies and therefore, IPs and EAs.

The static inverse dielectric function ($\epsilon_{\mathbf{G,G'}}^{-1}(\mathbf{q})$) is expanded in planewaves (with wavector~$\mathbf{G}$), and a cutoff 
($\epsilon_{\mathbf{G}}^{cut} = |\mathbf{q}+\mathbf{G}|^2/2$), where $\mathbf{q}$ is a wavevector. $\epsilon_{\mathbf{G,G'}}^{-1}(\mathbf{q})$ is 
constructed as a sum over unoccupied states~\cite{hybertsen_ab_1987}, which is truncated at a finite number of states, N$_\epsilon$, 
with energy E(N$_\epsilon$). The dielectric function is extended to finite frequency with the generalized plasmon-pole (GPP) model of Hybertsen and Louie~\cite{hybertsen_ab_1987}. 

For the purposes of analysis, we define the self-energy operator as a sum of Fock exchange, screened exchange, and the CH terms~\cite{hybertsen_electron_1986}.
The screened exchange term, $\Sigma_{\textrm{SX}}$, requires an explicit sum over just occupied states; however, it is implicitly dependent on $N_\epsilon$ through 
the dielectric function. The CH term, $\Sigma_{\textrm{CH}}$, involves a sum which in principle must span the full unoccupied subspace, but in practice is also 
truncated at finite number of unoccupied states $N_c$, with corresponding energy E($N_c$). For simplicity, we set $N_\epsilon$ equal to $N_c$, subtract the matrix 
elements of the Fock operator, $\Sigma_{\textrm{X}}$ (which is independent of $N_c$ and $\epsilon$), from $\Sigma_{\textrm{SX}}$, and study the convergence behavior 
of $\Sigma_{\textrm{SX-X}}$ and $\Sigma_{\textrm{CH}}$ terms with respect to $N_c$ and $\epsilon_{\mathbf{G}}^{cut}$. 

\section{Results}


\begin{figure*}[!h]
  \begin{center}
    \includegraphics[width=0.98\textwidth,clip]{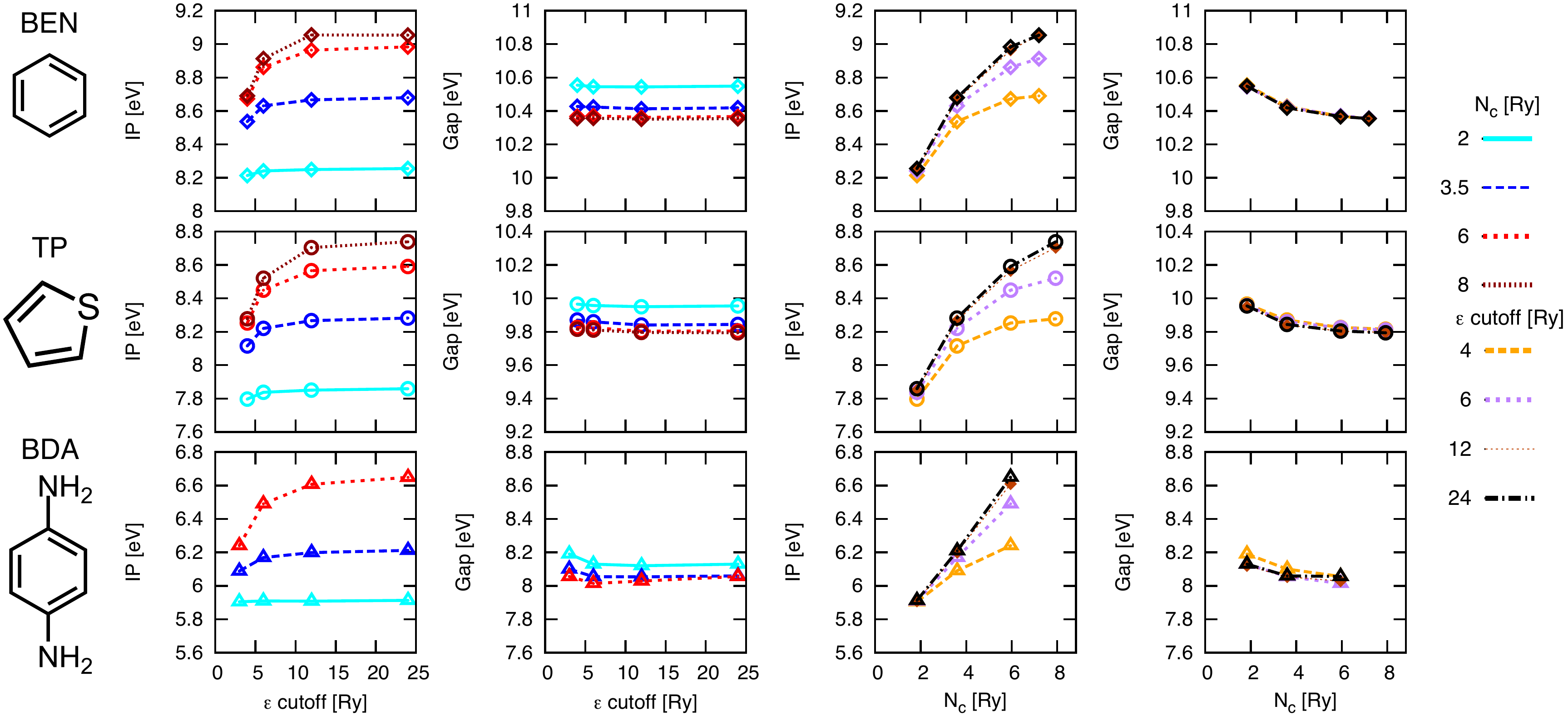}
  \end{center}
  \caption{\label{f:IE_convergence_epsilon} The $G_0W_0$ predicted IP and fundamental gap as a function of $\epsilon^{-1}$ cutoff and $N_c$.}
\end{figure*}

\subsection{Convergence of the dielectric matrix}

Fig.~\ref{f:IE_convergence_epsilon} summarizes our calculated IPs for BEN, TP, and BDA as a function of two parameters, $\epsilon_{\mathbf{G}}^{cut}$ and $N_c$. 
The IP is defined here such that a positive value indicates a bound electron. The IP increases significantly, by about 0.5 eV, as either parameter is 
increased (taken towards convergence) for all three molecules; in contrast, the fundamental gap, IP-EA, converges rapidly to within $0.1$ eV for $E(N_c) > 2$ Ry 
and $\epsilon_{\mathbf{G}}^{cut}$ $>4$ Ry. 

As noted previously~\cite{shih_quasiparticle_2010}, the interdependence of $N_c$ and $\epsilon_{\mathbf{G}}^{cut}$ can lead to a ``false convergence'' of the IP
 with respect to the dielectric cutoff at small fixed $N_c$. For all three molecules at $E(N_c) = 2$ Ry and $\epsilon_{\mathbf{G}}^{cut}$ $= 4$ Ry, the IP is apparently 
converged to within $0.1$ eV; however, if $ E(N_c)$ is increased to $6$ Ry, the IP varies by $0.3-0.4$ eV as $\epsilon_{\mathbf{G}}^{cut}$ is raised from $4$ to $24$ Ry. 
For all three molecules studied, this ``false convergence'' subsides for $E(N_c)\sim6$ Ry above the vacuum level (corresponding to $N_c$ $\sim3000$ for BEN and TP, 
and $\sim5000$ for BDA within our supercells); the computed IP is unaffected by an increase of $\epsilon_{\mathbf{G}}^{cut}$ for values greater than $12$ Ry for $E(N_c) \ge 6$ Ry. 
However, the IP is still quite sensitive to $N_c$, as we will discuss further below.

\begin{figure*}
\resizebox{0.9\textwidth}{!}{
\includegraphics{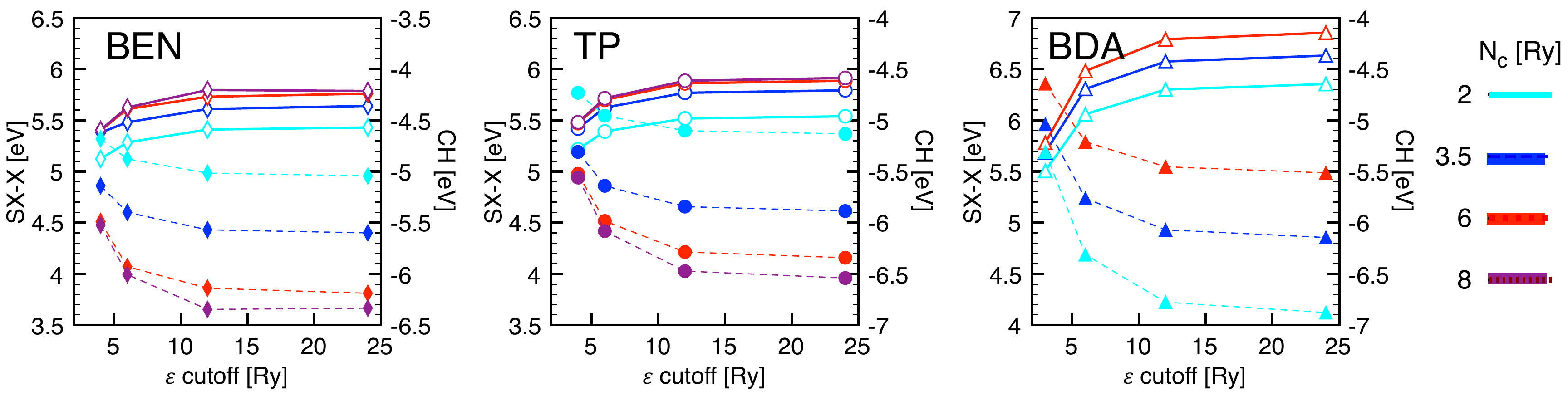}
}
\caption{For BEN, TP, and BDA HOMOs: $\Sigma_{\textrm{sx-x}}$ (solid lines) and $\Sigma_{\textrm{CH}}$ (dashed lines) as a function N$_c$ and $\epsilon_{\mathbf{G}}^{cut}$ for $E(N_C) = 25\textrm{, }50\textrm{, }80, \textrm{ and }105$ eV. The legend follows Fig. 1.}
\label{f:screened_exchange_coulomb_hole} 
\end{figure*}

For fixed $N_c$, both $\Sigma_{\textrm{SX-X}}$ and $\Sigma_{\textrm{CH}}$ also appear converged (to within $0.1$ eV) for  $\epsilon_{\mathbf{G}}^{cut}$ $\ge 12$ Ry, 
as shown in Fig.~\ref{f:screened_exchange_coulomb_hole} for the highest occupied molecular orbitals (HOMOs) of BEN, TP, and BDA. Interestingly, for low $N_c$ 
the variation of $\Sigma_{\textrm{SX-X}}$ and $\Sigma_{\textrm{CH}}$ with $\epsilon_{\mathbf{G}}^{cut}$ ranges from $2-100$ times larger than the corresponding variation of the IP. 
Thus, $\Sigma_{\textrm{SX-X}}$ and $\Sigma_{\textrm{CH}}$ are evidently less prone to ``false convergence'' at low $N_c$ than the IP. Since both $\Sigma_{\textrm{SX-X}}$
 and $\Sigma_{\textrm{CH}}$ depend on $\epsilon^{-1}$, but with opposite sign~\cite{hybertsen_ab_1987}, their sum (which determines the IP) is less sensitive to 
an underconverged dielectric function.

While Fig.~\ref{f:IE_convergence_epsilon} and Fig.~\ref{f:screened_exchange_coulomb_hole} suggest that an $\epsilon_{\mathbf{G}}^{cut}$ $\ge 12$ Ry is sufficient for a precision of $0.1$ eV or better in the IP for fixed $N_c$, they also highlight the fact that the self-energy corrections are more sensitive to $N_c$ when the high energy Fourier components of $\epsilon^{-1}$ are well-described. Fig.~\ref{f:IE_convergence_epsilon} shows a variation in IP of $>~1$ eV as E($N_c$) grows from $2$ Ry to greater than $6$ Ry. Fig.~\ref{f:screened_exchange_coulomb_hole} indicates that the $\Sigma_{\textrm{CH}}$ term is responsible for this variation, as $\Sigma_{\textrm{SX}}$ appears converged within $0.2$ eV for a dielectric matrix described with $E(N_c) \ge 6$ Ry. This implies that for the molecules and supercells under study, for $\epsilon_{\mathbf{G}}^{cut}$ $\ge 12$ Ry and $E(N_c) \ge 6$ Ry, the only remaining convergence issue in the calculation is the sum over the unoccupied subspace. We now discuss the different strategies for converging this sum. 

\subsection{Convergence of the Coulomb-hole term of the self-energy}

\begin{figure*}
\resizebox{0.9\textwidth}{!}{
	\includegraphics{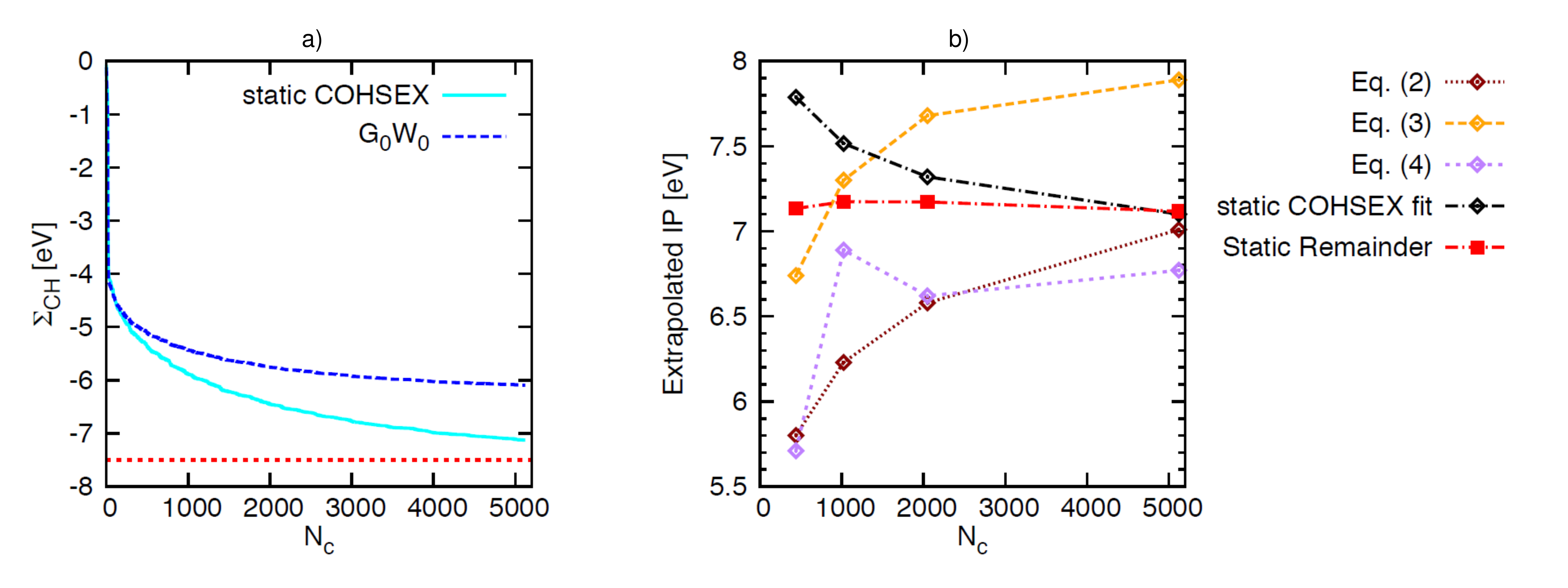}
}
\caption{\label{f:Fig3COH}  a) $\Sigma_{\textrm{CH}}$ as a function of number of bands for the BDA HOMO for both static COHSEX and $G_0W_0$. The static COHSEX result for $N_c \rightarrow \infty$ is indicated with a horizontal dotted line. b) The $G_0W_0$ IP with the CH term extrapolated to infinite $N_c$ using fitting techniques and the static remainder approach.}
\end{figure*}

The slow convergence of the $\Sigma_{\textrm{CH}}$ term, for a converged value of  $\epsilon_{\mathbf{G}}^{cut}$, with respect to $N_c$ can be seen in Fig.~\ref{f:Fig3COH}a for the BDA HOMO. $\Sigma_{\textrm{CH}}$ varies by more than $2$ eV for $N_c \in \left[ 500 ; 5000\right] $  and shows a finite slope of $10^{-4}$ eV/$N_c$ at $N_c = 5000$. Moreover, this same slow convergence behavior can be seen with a static CH and screened exchange method (static COHSEX) for which a full evaluation (shown as dashed line) does not require a sum of empty states~\cite{hedin_new_1965}. Comparison of our dynamic and static calculations suggests that the $N_c$ dependence of $\Sigma_{\textrm{CH}}$ comes from \emph{both} static and dynamical correlation terms. The static COHSEX CH term is still $0.2$ eV away from the exact solution at $N_c = 5000$, and has a different slope than the full dynamical $\Sigma_{\textrm{CH}}$. 

The slow convergence of $\Sigma_{\textrm{CH}}$ with $N_c$ has been addressed with different strategies in prior work~\cite{bruneval_accurate_2008,deslippe_coulomb_2011,kang_enhanced_2010,umari_gw_2010,giustino_gw_2010,berger_ab_2010,reining_elimination_1997}. Here, we examine three different approaches for extrapolating the CH term to infinite $N_c$ and examine their consequence for the IP: $i)$ fitting $\Sigma_{\textrm{CH}} \left(N_c \right)$ for a given orbital with an analytical form, and calculating its limit when $N_c \rightarrow \infty$ (see, e.g. ~\cite{kang_fit}); $ii)$ fitting the dynamical $\Sigma_{\textrm{CH}} \left(N_c \right)$ to a functional form determined from the corresponding static COHSEX term~\cite{tamblyn_electronic_2011}; and $iii)$ approximating the correction to the dynamical CH term based on completing the unoccupied subspace within the static COHSEX approximation, \emph{i.e.} the static remainder (SR) approach~\cite{deslippe_coulomb_2011}. 

Kang and Hybertsen applied a fitting scheme to $\Sigma_{\textrm{CH}}$ to obtain the valence band maximum of TiO$_{\textrm{2}}$ and found a $0.2$ eV range in predicted values for two different functional forms for the fit~\cite{kang_fit}. We take a similar approach and consider the following four functional forms for the dynamical $\Sigma_{\textrm{CH}} \left(N_c \right) $:

\begin{eqnarray}
	\Sigma_{\textrm{CH}}   \left(N_c \right) &\simeq& \alpha + \beta N_c^{-\frac{1}{\gamma}}, \label{Eq:Fit1}\\
	\Sigma_{\textrm{CH}}   \left(N_c \right) &\simeq& \alpha + \beta N_c^{-{1}}, \label{Eq:Fit2} \\
	\Sigma_{\textrm{CH}}   \left(N_c \right) &\simeq& \alpha + \beta N_c^{-\frac{1}{3}}, \label{Eq:Fit3}\\
	\Sigma_{\textrm{CH}}   \left(N_c \right) &\simeq& \alpha + \beta  \textrm{e}^{-\frac{N_c}{\gamma}}, \label{Eq:Fit2}
\end{eqnarray}

where $\alpha$, $\beta$, and $\gamma$ are fitting parameters. In practice, we find that good fits (P value $<$ 0.005) can be consistently obtained using any of these forms.  

We also fit the partial sum $\Sigma_{\textrm{CH}} (N_c)$ computed within static COHSEX such that $\alpha$ is the numerically exact closed form value of the static CH 
($\Sigma_{\textrm{CH}}^{\textrm{static}}(\infty)$). More precisely, the static CH term,  $\Sigma_{\textrm{CH}}^{\textrm{static}} \left(N_c \right) $, is fit to Eq.~\ref{Eq:Fit1}, with $\beta$ and $\gamma$ as fitting parameters. The dynamical $\Sigma_{\textrm{CH}} (N_c)$ is then fit to Eq.~\ref{Eq:Fit1}, with $\gamma$ fixed and $\alpha$ and $\beta$ as fitting parameters. Here, we are assuming that the same functional form describing the static $\Sigma_{\textrm{CH}}$ also describes the dynamical case. 

Lastly, we apply the SR correction defined in Ref.~\cite{deslippe_coulomb_2011} where 

\begin{equation}\Sigma_{\textrm{CH}} \left( N_c \rightarrow \infty \right) \simeq \Sigma_{\textrm{CH}} \left( N_c \right) + \frac{1}{2} \left[ \Sigma_{\textrm{CH}}^{\textrm{static}}(\infty) - \Sigma_{\textrm{CH}}^{\textrm{static}} \left( N_c \right) \right].
\end{equation}

In Fig.~\ref{f:Fig3COH}b), we report the computed IPs of BDA using all five extrapolation techniques described above. Because we are far from convergence in $N_c$, the fitting procedure $(i)$ is much less favorable than found by Ref.~\cite{kang_fit} both by its error with respect to experiments and its range of uncertainty: the assigned functional form can produce predicted IPs ranging from 5.8 to 7.2 eV. More importantly, the computed IP is very sensitive to the number of bands initially used. 
The best fit to the static COHSEX result for  $\Sigma_{\textrm{CH}}$, $(ii)$, results in IPs that monotonically increase with the number of bands used in the fit, and appears to be converging towards the SR result. 

The SR method gives the best results, with predicted IP values within $0.1$ eV for $N_c \in \left[ 500 ; 5000\right]$. The results of the SR method are particularly remarkable in the sense that when using this procedure, \emph{less} unoccupied states are needed to converge the CH term than the dielectric matrix (respectively $500$ and $5000$ for BDA).

\subsection{Comparison with experiment}

Table~\ref{tbl:withSR} shows the $G_0W_0$ IP and EA for BEN, TP, and BDA, along with experimental values. For all molecules, we use $E(N_c)= 6$ Ry and $\epsilon_{\mathbf{G}}^{cut}$ = $24$ Ry, and $\Sigma_{\textrm{{CH}}}$ is extrapolated to infinite number of bands via SR ~\cite{deslippe_coulomb_2011}. Our $G_0W_0$ results are in excellent agreement with experiment, within $0.2$ eV for IP of all three molecules and the EA of BEN. Our predictions agree well with previous planewave-based $G_0W_0$ studies~\cite{umari_gw_2010,umari_2009,samsonidze_simple_2011} for BEN, but differ somewhat quantitatively with with other $G_0W_0$ results obtained using localized basis sets for TP~\cite{blase_first-principles_2011} and BDA~\cite{strange_2011}.

\begin{table*}
\caption{$G_0W_0$ IP for BEN, TP, and BDA in eV. The calculations are performed with $\epsilon_{\mathbf{G}}^{cut}$ or $24$ Ry, with E(N$_c$) fixed at $6$ Ry, and the static remainder correction applied. }
\label{tbl:withSR}
\begin{tabular}{|c|c|c|c|}
\hline
 Molecule & BEN   &TP &BDA  \\
\hline
IP Theory &   9.4	&	9.0&	7.3	\\
\hline
IP experiment & 9.24~\cite{lias_ionization_2012}  &  8.86~\cite{lias_ionization_2012} & 7.34~\cite{streets_mesomeric_1972,cabelli_upe_1981}  \\
\hline
EA Theory &   -0.92	&	-0.94&	-0.90\\
\hline
EA experiment & -1.1~\cite{ben_ea}  &  -----~\cite{tp_ea} & -----  \\
\hline
\end{tabular}
\end{table*}

\section{Conclusions}

With use of unoccupied states that span $\sim6$ Ry in energy, an $\epsilon_{\mathbf{G}}^{cut}$ greater than or equal to $12$ Ry, and the static remainder approach to correct for the finite number of empty states in $\Sigma_{\textrm{CH}}$, we obtain converged values for the $G_0W_0$-calculated IP and EA of three organic molecules in the gas-phase. The predicted IPs and EAs agree to within $0.2$ eV with available experiment. Our results indicate that $G_0W_0$ provide quantitatively accurate addition/removal energies for small organic molecules. We find that a limiting step to these calculations is the large $\epsilon_{\mathbf{G}}^{cut}$ required for convergence. Thus, extrapolation techniques for $\epsilon_{\mathbf{G}}^{cut}$ will be increasingly valuable for describing larger systems, such as metal/organic molecule interfaces. 

\section{Acknowledgment}

The authors acknowledge fruitful discussions with J. Deslippe, G. Samsonidze, D. Strubbe (UC Berkeley), X. Blase (Institute Neel), and Leeor Kronik (Weizmann Institute).
Work at the Molecular Foundry was supported by the Office of Science, Office of Basic Energy Sciences, of the U.S. Department of Energy under Contract No. DEAC02-05CH11231. P.T.D.  was funded by the Helios Solar Energy Research Center, which is supported by the Director, Office of Science, Office of Basic Energy Sciences of the U.S. Department of Energy under Contract No. DE-AC02-05CH11231.Computer simulations were performed using NERSC. S.S. and P.D acknowledge funding from the National Science Foundation through the Network for Computational Nanotechnology (NCN) and I.T. acknowledges financial support from NSERC.

\bibliography{GW_Europhysics}

\end{document}